\documentclass[twocolumn,showpacs,preprintnumbers,amsmath,amssymb
              ,prb,eqsecnum]{revtex4}

\usepackage{graphicx}
\usepackage{dcolumn}
\usepackage{bm}

\begin{document}

\preprint{two-leg L-edge}

\title{Theory of $L$-edge resonant inelastic x-ray 
scattering for magnetic excitations in two-leg 
spin ladder}

\author{Tatsuya Nagao}
\affiliation{%
Faculty of Engineering, Gunma University, Kiryu, Gunma 376-8515, Japan}

\author{Jun-ichi Igarashi}%
\affiliation{%
Faculty of Science, Ibaraki University, Mito, Ibaraki 310-8512, Japan}

\date{\today}

\begin{abstract}

We study the magnetic excitation spectra of Cu 
$L_{2,3}$-edge resonant inelastic x-ray scattering 
(RIXS) from the spin-liquid
ground state in two-leg spin ladder cuprates.
Applying the projection method developed by the present authors,
we have derived the formulas of the magnetic RIXS spectra, which are 
expressed by one- and two-spin correlation functions in the polarization 
changing and preserving channels, respectively.
The one-spin correlation function
includes both one- and two-triplon excitations
and they are picked up separately by choosing rung wave vector $-q_a=\pi$ and $0$, respectively.
An application to Sr$_{14}$Cu$_{24}$O$_{41}$
reveals that the calculated RIXS spectrum 
captures well the dispersive behavior shown by 
the lower boundary of two-triplon continuum. 
By adjusting the geometrical 
configuration of the measurement, one-triplon
dispersion around the zone center could be 
detectable in the RIXS experiment.
The observed weak intensity in the higher 
energy region might be attributed to the 
two-spin correlation function, which could be detected
more clearly at the $M_{3}$-edge RIXS spectra. 
\end{abstract}

\pacs{78.70.Ck, 72.10.Di, 78.20.Bh, 74.72.Cj} 
\maketitle

\section{\label{sect.1}Introduction}
Resonant inelastic x-ray scattering (RIXS)
provides us with one of the rare opportunities to 
investigate various excitations in solids 
including charge, orbital, and spin degrees of
freedoms.\cite{Ament2011}
Among them, extracting the magnetic excitation
by RIXS is rather challenging because 
to distinguish its intensity from
other contributions such as the elastic signal and phonon excitations is hard to achieve due to 
the small energy transfer
below 1 eV associated with the magnetic excitation.
However, it is worthwhile to pursue 
potential of RIXS as a promising 
probe for magnetic excitation, since it can 
survey a wide range of momentum transfer
in the Brillouin zone and requires only small sample volume. These features convince us RIXS 
to be a choice complementary to the conventional 
inelastic neutron scattering (INS).

Recently, progress in the experimental 
technique of RIXS enables us to carry out 
higher energy resolution 
measurement,\cite{Ghiringhelli2004,Ghiringhelli2006} 
for instance, in several transition
metal oxides like
the cuprates.\cite{Hill2008,Braicovich2009,
Bisogni2009,Schlappa2009}
Hill \textit{et al}. have succeeded in detecting the
magnetic excitation spectra peaked around
500 meV energy region at the Cu $K$-edge
in La$_{2}$CuO$_{4}$.\cite{Hill2008}
The following theories have revealed that the
obtained signals are attributed to the two-magnon
excitation from the antiferromagnetic (AFM) 
ground state brought about by the presence of 
the core hole potential during the intermediate
states.\cite{vdBrink2007,Nagao2007,Forte2008}
Since then, plenty of measurements have been reported 
out at the Cu $K$-edge as well as 
Cu $L_{3}$-edge.\cite{Ellis2010,Bisogni2010,
Guarise2010,Braicovich2010,Braicovich2010.2}

When we turn our attention to the $L$-edge
RIXS, the situation becomes a little complicated.
In the Cu $L$-edge RIXS process, the transition
is between the $2p$ core and $3d$ states.
The photo-excited electron eliminates the
$3d$ hole leading the $3d$ state to the closed shell.
The absence of the spin degree of freedom
at the core-hole site is similar to the situation 
of the non-magnetic impurity problem
in the spin system.\cite{Tonegawa1968,Wan1993}
Constructing a relevant theory to handle 
such difficult situation is 
challenging but attracting.  
Within the fast collision approximation,
the momentum and polarization dependences
of the magnetic excitation spectra at the Cu 
$L_{2,3}$-edges were 
investigated.\cite{Luo1993,Veenendaal2006,Ament2009}
In another attempts, resonant energy dependence of 
the RIXS spectra is described by the scattering 
operator inferred
from elastic scattering.\cite{Hannon1988,Haverkort2010}

In order to extract the magnetic excitation from
the closed $3d$ state at the core hole site 
in the intermediate state,
we have developed an effective theory to investigate the
magnetic RIXS spectrum at the Cu $L_{2,3}$ 
edges.\cite{Igarashi2012.2D,Igarashi2012.1D} 
The theory treats the polarization and 
energy dependences of the
spectrum faithfully, projecting the final states
onto the possible spin excited states, from which 
the RIXS spectra are expressed as the form of
the spin correlation functions.
They consist of the two-spin
as well as one-spin correlation functions.
The latter is also found in the theory of INS,
but the former is specific to RIXS.
Even for the one-spin correlation function,
its transferred momentum dependence turns out to be 
completely different from the one in the
INS theory reflecting the inclusion of the 
non-local magnetic excitation.

We have applied our theory to two-dimensional system 
having the AFM ground state and 
confirmed the quantitative effectiveness of
the theory.\cite{Igarashi2012.2D} 
It has reproduced well the 
experimental data reported by Guarise \textit{et al}. in 
Sr$_2$CuO$_{2}$Cl$_2$.\cite{Guarise2010}
Contributions of the multi-magnon excitations
have been evaluated too.
Then, we have applied our theory to one-dimensional Heisenberg
chain having the spin-singlet 
ground state\cite{Igarashi2012.1D} that preserves rotational
invariance in the spin space.
The RIXS spectrum, which can cover the magnetic excitations generated 
not only at the core-hole site but
also around the core-hole site, has been derived in a way that
manifests the rotational invariance.
The results have been consistent with those
obtained by other theoretical 
approaches.\cite{Klauser2011,Kourtis2012,Forte2011} 
Our result also suggests
that contribution from the two-spin correlation function is detectable at the higher energy
region in the $\sigma$-polarization measurement.
Unfortunately, at that time, no experimental data 
was available to compare with our results.
 
In this work, we extend our theory to exploit 
the Cu $L_{2,3}$-edge magnetic excitation spectrum in  
RIXS for two-leg spin ladder cuprate where the ground
state is the spin-liquid retaining the
spin rotational invariance. 
The low energy sector of the spin excitations
are known as one- and two-triplon 
excitations.\cite{Schmidt2005} 
Our analysis reveals that the RIXS spectra can
trace the one- and two-triplon dispersions as a function
of the transferred momentum through the one-spin 
correlation function. 

\begin{figure}
\includegraphics[width=8.0cm]{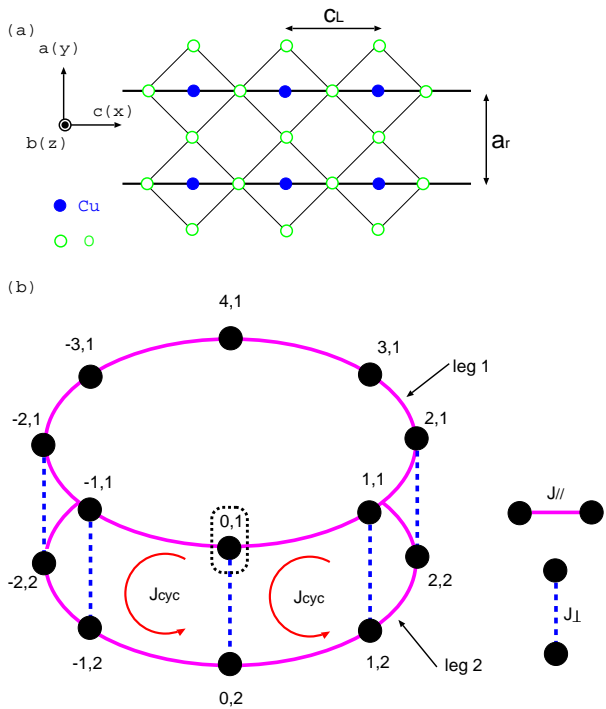}%
\caption{(Color online)\label{fig.two-leg}
Schematic sketch of the material and model.
(a) Cu-O network of ladder in Sr$_{14}$Cu$_{24}$O$_{41}$.
Filled (blue) and open (green) circles denote
Cu and O sites, respectively.
(b) Model of two-leg ladder $2 \times 8$ spins used to evaluate 
$f_{\mu}^{(1)}(\omega_i)$'s 
and $f_{\nu}^{(2)}(\omega_i)$'s.
The periodic boundary conditions are adopted.
The spin at site 0 on leg 1 (encircled by dotted line)
is annihilated in the intermediate state.}
\end{figure}

Then, we apply our theory to interpret the experimental
results reported by Schlappa \textit{et al}.
in Sr$_{14}$Cu$_{24}$O$_{41}$.\cite{Schlappa2009}
This material is considered an experimental
realization of a two-leg ladder 
structure (Fig. \ref{fig.two-leg} (a)),\cite{McCarron1988}
and has attracted of particular interest 
since Ca doped
systems Sr$_{14-x}$Ca$_{x}$Cu$_{24}$O$_{41}$ have exhibited
superconducting state under pressure for $x$
ranging from 11.5 to 13.6.\cite{Uehara1996}
In the RIXS measurement at the Cu $L_3$-edge (930.6 eV),
Schlappa \textit{et al}. observed the dispersive signals attributed to the
lower boundary of two-triplon excitation continuum.
Our theory succeeds in reproducing
the momentum dependence of the RIXS spectral profile. 
In addition, we find one-triplon dispersion can be
detected by rotating the sample around the $b$-axis.
The observed weak intensity in the higher 
energy region might be the contribution from 
the two-spin correlation function, mainly originated
from two-triplon excitations with total spin $S=0$. 
We argue that the contribution from the pure two-spin 
correlation function is distinguishable more clearly
at the $M_{3}$-edge measurement.

The present paper is organized as follows.
In Sec. \ref{sect.2}, we extend
the theory developed in our previous papers
aiming at the application to the Cu $L_{2,3}$ edges
in two-leg ladder systems.
The RIXS spectra are expressed 
in terms of spin-correlation functions. 
In Sec. \ref{sect.3}, the amplitudes leading to 
the spin-correlation functions are evaluated on a finite-size two-leg ladder cluster. An application to 
Sr$_{14}$Cu$_{24}$O$_{41}$ 
is shown in Sec. \ref{sect.4}. 
Section \ref{sect.5} is devoted to summary and discussion.

\section{Theoretical framework\label{sect.2}}
\subsection{\label{sect.2.C}Initial state 
and magnetic excitations}
For the purpose of application to the two-leg spin ladder
system Sr$_{14}$Cu$_{24}$O$_{41}$, we consider 
the system is at half-filling and 
is \textit{undoped}\cite{Com0} with each Cu atom having
one hole per site in the $x^2-y^2$ 
orbital.\cite{Nucker2000,Gozar2001}
The $x$ and $y$ axes are defined along the Cu-O bonds
parallel to the crystallographic $c$ and $a$ axes, respectively, 
while the $z$ along the $b$ axis.
Here, the $a$ and $c$ axes are along the rung and 
leg directions of the ladders.
The ground state and the low-energy spin excitations
are described by the $S=1/2$ antiferromagnetic Heisenberg 
Hamiltonian with an additional four-spin 
exchange terms,\cite{MullerHartmann2002}
\begin{eqnarray}
 H_{\rm mag}&=&J_{\parallel} \sum_{i}\sum_{\tau=1,2} 
    \textbf{S}_{i,\tau}\cdot \textbf{S}_{i+1,\tau}
+J_{\perp} \sum_{i} 
  \textbf{S}_{i,\tau}\cdot \textbf{S}_{i,\overline{\tau}} 
\nonumber \\
&+& J_{\textrm{cyc}}
\sum_{\textrm{plaquette}}
(\textbf{S}_{i,\tau}\cdot \textbf{S}_{i+1,\tau})
(\textbf{S}_{i,\overline{\tau}}\cdot \textbf{S}_{i+1,\overline{\tau}})
\nonumber \\
&+&J_{\textrm{cyc}}
\sum_{\textrm{plaquette}}
(\textbf{S}_{i,\tau}\cdot \textbf{S}_{i,\overline{\tau}})
(\textbf{S}_{i+1,\tau}\cdot \textbf{S}_{i+1,\overline{\tau}})
\nonumber \\
&-& J_{\textrm{cyc}}
\sum_{\textrm{plaquette}}
(\textbf{S}_{i,\tau}\cdot \textbf{S}_{i+1,\overline{\tau}})
(\textbf{S}_{i+1,\tau}\cdot \textbf{S}_{i,\overline{\tau}}),
\end{eqnarray}
where the index $i$ refers to the rungs.
The symbol $\tau$ $\in \{ 1,2 \}$ discriminates legs, and 
$\overline{\tau}$ denotes $2$ for $\tau=1$, 
and vice versa.
The exchange coupling constants along the
legs and rungs are denoted as $J_{\parallel}$ 
and $J_{\perp}$, respectively (Fig. \ref{fig.two-leg} (b)). 
In addition, four-spin coupling $J_{\textrm{cyc}}$ is included.  

We briefly summarize the ground state
and low energy
sector of the spin excitations of $H_{\textrm{mag}}$.
The ground state is known to be the gapped
spin-liquid due to the
quantum fluctuation.
\cite{Dagotto1988,Dagotto1992,Greven1996,
Dagotto1996,Shelton1996}
Unlike the AFM ordered state,
the spin-liquid state preserves the rotational invariance.
First, let us assume $J_{\parallel}=J_{\textrm{cyc}}=0$.
Since each rung is independent, the ground state
is constructed by $N/2$ pairs of rung singlet 
where $N$ denotes the number of spins. This state
is called as \textit{rung singlet}.
The lowest energy excited state is realized
by changing one of the singlets into triplet with the
excitation energy, or equivalently a spin gap, $J_{\perp}$.
As the number of triplets increases, the excited
states are called as one-, two-, three-triplon
excitations, and so on.

When $J_{\parallel}$ is turned on, this kind of simple
picture may be invalid. However, it is known that
even when $J_{\parallel} \rightarrow \infty$,
the ground state is adiabatically connected to
the \textit{rung singlet}.\cite{Barnes1993} 
Thus, classifying the magnetic excitation
by the number of triplons gives a good description
and the spin gap remains finite.\cite{Dagotto1988,Dagotto1992,Greven1996,
Dagotto1996,Shelton1996,Schmidt2005}
The presence of $J_{\textrm{cyc}}$ does not alter
the nature of the ground and low energy excited states
qualitatively for
$J_{\textrm{cyc}}/J_{\perp} \sim 0.25$, although
complication sets in when the ratio grows, 
which is beyond the present interest.\cite{Hikihara2003} 

\subsection{RIXS spectra}
We extend our RIXS theory developed for
spin-singlet one-dimensional Heisenberg 
chain\cite{Igarashi2012.1D}
to quasi one-dimensional two-leg spin ladder 
system showing the spin-liquid ground state.
In the following, a concise version of the
explanation of the theory is displayed, 
relegating a detail to 
ref. \onlinecite{Igarashi2012.1D}.

In the electric dipole ($E$1) transition at the 
transition-metal $L_{2,3}$-edge, the transition
progresses between the $2p$-core state and the $3d$ state.
This process, described by the electron-photon interaction
Hamiltonian $H_{\textrm{int}}$, 
is mediated by absorbing the incident
photon with wave vector $\textbf{q}_i$ and energy
$\omega_i$, and then, emitting the scattered photon 
with wave vector $\textbf{q}_f$ and energy
$\omega_f$.
The RIXS spectra may be expressed by the second-order
$E$1 allowed 
process:\cite{Igarashi2012.2D,Igarashi2012.1D}
\begin{eqnarray}
 W(q_f \alpha_f;
   q_i \alpha_i)
 &=& 2\pi\sum_{f'}\left|\sum_{n}
  \frac{\langle \Phi_{f'}|H_{\rm int}|n\rangle
        \langle n|H_{\rm int}|\Phi_i\rangle}
       {E_g+\omega_i-E_n} \right|^2 \nonumber\\
 &\times&\delta(E_g+\omega_i-E_{f'}-\omega_f),
\label{eq.optical}
\end{eqnarray} 
with $q_i\equiv({\bf q}_i,\omega_i)$, $q_f\equiv({\bf q}_f,\omega_f)$, 
$|\Phi_i\rangle = c_{{\bf q}_i \alpha_i}^{\dagger}|g\rangle$,
$|\Phi_{f'}\rangle=c_{{\bf q}_f \alpha_f}|f'\rangle$,
where $|g\rangle$ and $|f'\rangle$ represent the ground state and
excited states of the matter with energy $E_g$ and $E_{f'}$, respectively.
The polarization directions of the incident
and scattered photons are $\alpha_{i}$
and $\alpha_{f}$, respectively.
The annihilation (creation) operator of photon 
with momentum ${\bf q}$ and polarization 
$\alpha$ is denoted
as $c_{{\bf q}\alpha}$
($c_{{\bf q}\alpha}^{\dagger}$).
The intermediate state is denoted as $|n \rangle$ with 
energy $E_{n}$ in the presence of the core-hole.
Since $E_n$ includes the core-hole energy 
$\epsilon_{\textrm{core}}$, we express
$E_n$$=\epsilon_{\textrm{core}}- i \Gamma$$+\epsilon_{n}$
where $\Gamma$ stands for the lifetime broadening width
of the core-hole and $\epsilon_n$ is the energy
of the spin part in the intermediate state.
The $\epsilon_{n}$ will be evaluated by the Hamiltonian 
$H_{\textrm{mag}}'$,
constructed from $H_{\textrm{mag}}$ by eliminating the
spin degree of freedom at the central core-hole site.

In the intermediate state, the spin degree of freedom is lost
at the core-hole site. The final state experienced such intermediate state 
may be expressed by ${\bf S}_{0,1}|g\rangle$, ${\bf S}_{0,2}|g\rangle$,
${\bf S}_{1,1}|g\rangle$, $\cdots$ in the polarization changing channel,
and ${\bf S}_{0,1}\cdot{\bf S}_{1,1}|g\rangle$,
${\bf S}_{0,1}\cdot{\bf S}_{0,2}|g\rangle$, $\cdots$ in the polarization 
preserving channel, where suffix $(0,1)$ indicates the core-hole site.
Paying attention to the non-orthogonality of these state, 
we project the final state on these state in the same way as carried out
in our previous study on the one-dimensional system. We obtain
\begin{eqnarray}
 & & \sum_{n}
  \frac{H_{\rm int}|n\rangle
        \langle n|H_{\rm int}|\Phi_i\rangle}
       {E_g+\omega_i-\epsilon_{\textrm{core}} 
       - \epsilon_{n} + i \Gamma} \nonumber \\
    &=& \left(-\frac{i}{15}\right)
    \mbox{\boldmath{$\alpha$}}_f\times
    \mbox{\boldmath{$\alpha$}}_i\cdot 
     [f_{1}^{(1)}(\omega_i){\bf S}_{0,1} \nonumber \\
   &&+f_{2}^{(1)}(\omega_i)
     ({\bf S}_{1,1}+{\bf S}_{-1,1}) 
     +f_{3}^{(1)}(\omega_i){\bf S}_{0,2} \nonumber \\
   &&+f_{4}^{(1)}(\omega_i)
     ({\bf S}_{1,2}+{\bf S}_{-1,2})]|g\rangle, \nonumber \\
   &+& \frac{2}{15} \mbox{\boldmath{$\alpha$}}_{f\perp}\cdot
   \mbox{\boldmath{$\alpha$}}_{i\perp} 
   \left[f_{1}^{(2)}(\omega_i)+
     f_{2}^{(2)}(\omega_i){\bf S}_{0,1}
\cdot({\bf S}_{1,1}+{\bf S}_{-1,1}) \right. \nonumber\\
  &&+ \left. 
f_{3}^{(2)}(\omega_i){\bf S}_{0,1}\cdot{\bf S}_{0,2}
+ f_{4}^{(2)}(\omega_i){\bf S}_{0,1} 
\cdot({\bf S}_{1,2}+{\bf S}_{-1,2}) 
     \right]|g\rangle,
\label{eq.two} \nonumber \\
\end{eqnarray}
where $\mbox{\boldmath{$\alpha$}}_{i\perp}$ and 
$\mbox{\boldmath{$\alpha$}}_{f\perp}$
stand for the polarization vectors of the
incident and scattered photons, respectively,
projected onto the $xy$ plane.
The $f_{\mu}^{(n)}(\omega_i)$'s are the coefficients
to be determined. They could be accurately
evaluated in a system having rather small size,
since the relevant excited states are restricted around the core-hole site.

Accordingly, the RIXS spectra for the polarizations 
$\mbox{\boldmath{$\alpha$}}_{i(f)}
=(\alpha_{i(f)}^{x},\alpha_{i(f)}^{y},\alpha_{i(f)}^{z})$ 
are expressed as
\begin{eqnarray}
 W(q_f \mbox{\boldmath{$\alpha$}}_f;
   q_i \mbox{\boldmath{$\alpha$}}_i)
&\propto&
\left(\frac{\alpha_{f}^{x}\alpha_{i}^{y}
      -\alpha_{f}^{y}\alpha_{i}^{x}}{15}\right)^2
                  Y^{(1)}(\omega_i;q_c,q_a,\omega) 
            \nonumber \\
&+& 
\left[\frac{2\left(\alpha_{f}^{x}\alpha_{i}^{x}+\alpha_{f}^{y}
\alpha_{i}^{y}\right)}{15}\right]^2
                  Y^{(2)}(\omega_i;q_c,q_a,\omega), 
\nonumber \\
\label{eq.rixs.both}
\end{eqnarray}
where the first and second terms 
represent the contributions from 
the polarization changing and preserving channels, respectively. 
The $Y^{(1)}(\omega_i;q_c,q_a,\omega)$ and 
$Y^{(2)}(\omega_i;q_c,q_a,\omega)$ are Fourier transforms of the one-spin 
and two-spin correlation functions defined by
\begin{eqnarray}
& & Y^{(1)}(\omega_i;q_c,q_a,\omega) \nonumber \\
&=&
\int\langle Z^{(1)\dagger}(\omega_i;q_c,q_a,t) 
  Z^{(1)}(\omega_i;q_c,q_a,0)\rangle 
  {\rm e}^{i\omega t}{\rm d}t, 
\label{eq.y1} \\
&& Y^{(2)}(\omega_i;q_c,q_a,\omega) \nonumber \\
&=&
\int\langle Z^{(2)\dagger}(\omega_i;q_c,q_a,t) 
  Z^{(2)}(\omega_i;q_c,q_a,0)\rangle 
  {\rm e}^{i\omega t}{\rm d}t,
\label{eq.y2} 
\end{eqnarray}
with
\begin{eqnarray}
&& Z^{(1)}(\omega_i;q_c,q_a) \nonumber \\
&=&\sum_{j,\tau}
 {\rm e}^{-iq_cr_{j}-i q_a(\tau-1)a_r} \nonumber \\
&\times&
  [f_{1}^{(1)}(\omega_i) S_{j,\tau}^{z}
    +f_{2}^{(1)}(\omega_i)(S_{j+1,\tau}^{z}+S_{j-1,\tau}^{z})
 \nonumber \\
&+& f_{3}^{(1)}(\omega_i)S_{j,\overline{\tau}}^{z}
+f_{4}^{(1)}(\omega_i)(S_{j+1,\overline{\tau}}^{z}
                      +S_{j-1,\overline{\tau}}^{z}) ],
\label {eq.Z10} \\
&& Z^{(2)}(\omega_i;q_c,q_a) \nonumber \\
&=&
\sum_{j,\tau}{\rm e}^{-iq_cr_{j}+q_a(\tau-1)a_r} \nonumber \\
&\times &
   \left[f_{2}^{(2)}(\omega_i)
    ({\bf S}_{j+1,\tau}+{\bf S}_{j-1,\tau})
    + f_{3}^{(2)}(\omega_i){\bf S}_{j,\overline{\tau}}
    \nonumber \right. \\
   &+& \left. f_{4}^{(2)}(\omega_i)
   ({\bf S}_{j+1,\overline{\tau}}
   +{\bf S}_{j-1,\overline{\tau}})
     \right] \cdot {\bf S}_{j,\tau}. \label{eq.Z20}
\end{eqnarray}
The coordinate of the site at the $j$-th rung
and $\tau$-th leg is denoted as $(r_j,(\tau-1)a_r)$ in
the $ca$ plane.
The symbols $q_c$ and $q_a$ represent the wave numbers 
along the leg and rung directions, respectively.
They are the corresponding components of vector 
$\textbf{q}_{\perp}$,
which is the projection of the scattering vector
$\textbf{q}=\textbf{q}_i-\textbf{q}_f$ onto the $ca$-plane.
In the two-leg ladder configuration, the latter
takes two relevant values $0$ and $\pi/a_r$.
Hereafter, the momenta $q_c$ and $q_a$ are measured in units
of $1/c_L$ and $1/a_r$, respectively, when their
numerical values are mentioned.

Note that in the above expressions (\ref{eq.Z10}) and 
(\ref{eq.Z20}), the magnetic excitations included are
those 
at the neighboring sites within the two adjacent plaquettes linked 
to the central site as well as at the central core-hole site.
Notice also that if the fast collision
approximation is adopted, 
$Z^{(1)}(\omega_i;q_c,q_a)$ and 
$Z^{(2)}(\omega_i;q_c,q_a)$ become
$f_{1}^{(1)}(\omega_i) \sum_{j,\tau}
 {\rm e}^{-iq_cr_{j}-i q_a(\tau-1)a_r} 
  S_{j,\tau}^{z}$ and $0$, respectively, 
meaning that only the excitation
at the core-hole site is relevant to the former
while the latter vanishes.

The first term of Eq.~(\ref{eq.rixs.both}) gives the spectral shape 
as a function of $\omega$ similar to the conventional
one-spin correlation function familiar to INS theory;
the presence of $f_2^{(1)}(\omega_i)$
and $f_4^{(1)}(\omega_i)$ modifies the $q_c$ 
dependence of the spectral intensity.
The complicated $q_c$ dependence is a
direct consequence of the inclusion 
of non-local spin excitation around
the core-hole site as seen from Eq. (\ref{eq.Z10}),
which is missing in the fast collision approximation.
The second term of Eq.~(\ref{eq.rixs.both}) gives
the spectral shape arising from the non-local exchange type
excitations occurred around the core-hole site.
This type of contribution is missing in the
INS spectra and appears only when the theory
covers $\omega$ dependence beyond the fast
collision approximation.

We end this section with the explanation how
one- and two-triplon excitations are
manifested in the one- and two-spin
correlation functions.
For simplicity, we do not mention $n$-triplon
excitation with $n$ more than three
because $n=1$ and $2$ dominate quantitatively.
From the definitions Eqs. (\ref{eq.Z10})
and (\ref{eq.Z20}), one- and two-spin 
correlation functions are associated with
total spin $S=1$ and spin-conserving excitations,
respectively. Since one- and two-triplon
excitations are total spin $S=1$ and $S=0,1,2$,
respectively, one-spin correlation function
includes both one- and two-triplon excitations
with $S=1$ while two-spin correlation function
includes two-triplon excitation alone.

Note that, in the undoped ladder system, 
multi-triplon contributions
with different parity do not mix because
the system is invariant with respect to 
reflection about the centerline of the 
ladder.\cite{Schmidt2005}
Therefore, one- and two-triplon contributions
involved in the one-spin correlation 
function $Y^{(1)}(\omega_i;q_c,q_a,\omega)$
can be separated.
That is, one- and two-triplon contributions are
found in the $q_a=\pi$ and $q_a=0$ modes,
respectively, in which the
spin excitations in leg 1 and 2 are summed up in
anti-phase, and in phase, respectively.
 
\section{Numerical results\label{sect.3}}

In order to evaluate $f_{\mu}^{(1)}(\omega_i)$'s
and $f_{\nu}^{(2)}(\omega_i)$'s, we must
prepare the eigenstates and the corresponding
energies of $H_{\textrm{mag}}$
and $H_{\textrm{mag}}'$, the spin Hamiltonian in the intermediate state.
In addition, the incident photon energy 
$\omega_i$ should be specified, which is chosen as the
peak position of the absorption coefficient.
In the following, we shall describe how these
preparations are made.

\subsection{Eigenstates of $H_{\textrm{mag}}$
and $H_{\textrm{mag}}'$}
We consider a system consisting of $2 \times 8$ 
spins of $S=1/2$ 
with periodic boundary conditions for the initial and final states, 
as shown in Fig.~\ref{fig.two-leg} (b). 
The exchange couplings are chosen here 
as $J_{\parallel}=186$ meV, 
$J_{\perp}=124$ meV, and $J_{\textrm{cyc}}=31$ meV.\cite{Gozar2001,Nunner2002,Notbohm2007,Schmidt2005.2}
Representing $H_{\rm mag}$ by a matrix
of $12870\times 12870$ dimensions in the subspace of the $z '$ component of
the total spin $S_{\rm tot}^{z '}=0$, we diagonalize the Hamiltonian matrix.
We obtain the ground state energy as $\epsilon_g/(NJ_{\perp})=-0.733$.
The $H_{\textrm{mag}}'$ is obtained from $H_{\textrm{mag}}$ by eliminating
the spin at the core-hole site.
Therefore $H_{\textrm{mag}}'$ consists of 15 spins, and
may be represented by a matrix with 
$6435\times 6435$ dimensions in the subspace of $S_{\rm tot}^{z '}=\pm 1/2$.

\subsection{Absorption coefficient}

\begin{figure}
\includegraphics[width=8.0cm]{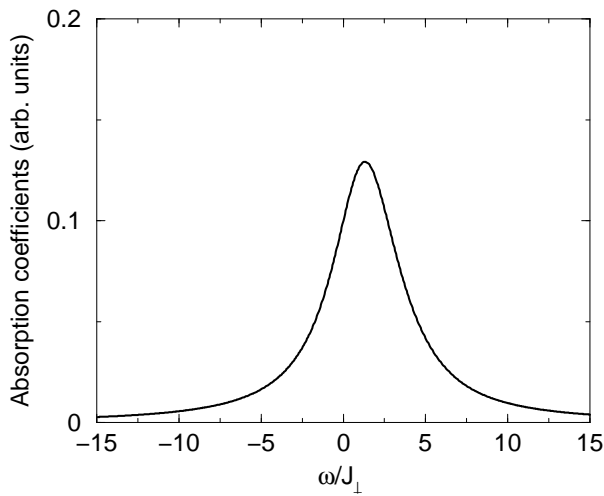}%
\caption{\label{fig.abs}
Absorption coefficients $A(\omega_i)$ as a function of photon energy $\omega_i$.
$\Gamma/J_{\perp}=2.4$. The origin of energy is set to correspond to
$\omega_i=\epsilon_{\textrm{core}}$. 
}
\end{figure}

By substituting the eigenvalues and the eigenstates evaluated on finite-size cluster
into Eq.~(A.1) of ref. \onlinecite{Igarashi2012.1D}, 
we obtain the $L_{2,3}$-absorption coefficient 
$A_j(\omega_i)$ ($j=3/2$ or $j=1/2$).
Figure \ref{fig.abs} shows the calculated $A_{j}(\omega_i)$
as a function of the incident photon energy. 
The origin of photon energy is set to be $\omega_i=\epsilon_{\textrm{core}}$. The $\epsilon_{\textrm{core}}$ 
stands for the energy required 
to create a $2p$ core-hole in the multiplet
$j=1/2$ or $3/2$ of the $(3d)^{10}$-configuration.
For Sr$_{14}$Cu$_{24}$O$_{41}$, we take 
$\Gamma/J_{\perp}=2.4$, since a typical value of 
$\Gamma \sim 300$ meV for Cu $L_3$-edge and 
$J_{\perp}=124$ meV.
The calculated curve turns out to be very close to the Lorentzian shape.
The peak position is slightly shifted from 
$\omega_i=\epsilon_{\textrm{core}}$, instead, it is 
around $\omega_i=\omega_i^0\simeq\epsilon_{\textrm{core}}
+1.365J_{\perp}$ for $\Gamma/J_{\perp}=2.4$.

\subsection{Evaluation of the coefficients}
\begin{table*}
\caption{\label{table.3}
Coefficients $f_{\mu}^{(1)}(\omega_i^0)$'s and $f_{\nu}^{(2)}(\omega_i^0)$'s 
in units of $1/J_{\perp}$ for different values 
of $\Gamma/J_{\perp}$
evaluated for $2 \times 8$ spin ladder.
}
\begin{ruledtabular}
\begin{tabular}{ccccc}
  $\Gamma/J_{\perp}$ & $f_{1}^{(1)}(\omega_i^0)$ & 
               $f_{2}^{(1)}(\omega_i^0)$ &
               $f_{3}^{(1)}(\omega_i^0)$ & 
               $f_{4}^{(1)}(\omega_i^0)$ \\
\hline
       $2.4$ & $(0.034,-0.849)$ & $(0.052,-0.037)$ &
               $(0.029,-0.026)$ & $(0.035,-0.026)$ \\
       $1.6$ & $(0.063,-1.285)$ & $(0.084,-0.085)$ &
               $(0.045,-0.055)$ & $(0.056,-0.060)$ \\
       $1.0$ & $(0.019,-2.080)$ & $(0.110,-0.195)$ &
               $(0.056,-0.116)$ & $(0.073,-0.134)$ \\
\hline
             &                  &                       \\
\hline
  $\Gamma/J_{\perp}$ & & 
               $f_{2}^{(2)}(\omega_i^0)$ & 
               $f_{3}^{(2)}(\omega_i^0)$ &
               $f_{4}^{(2)}(\omega_i^0)$ \\
\hline
       $2.4$ & & $(0.145,-0.108)$ & $(0.076,-0.060)$ & $(0.001,0.004)$ \\
       $1.6$ & & $(0.231,-0.246)$ & $(0.120,-0.113)$ & $(0.003,0.004)$ \\
       $1.0$ & & $(0.299,-0.550)$ & $(0.154,-0.292)$ & $(0.007,0.003)$ \\
\end{tabular}
\end{ruledtabular}
\end{table*}

By using the eigenvalues and eigenfunctions on 
a two-leg ladder of $2 \times 8$ spins, we calculate
the coefficients for $\omega_i=\omega_i^{0}$.
Table \ref{table.3} lists the calculated values.
For $f_{\mu}^{(1)}$'s, 
$|f_{2}^{(1)}|$ and $|f_{3}^{(1)}|$ are rather
smaller than $|f_{1}^{(1)}|$ with $\Gamma/J_{\perp}=2.4$,
while they become larger with $\Gamma/J_{\perp}$
increasing. 
It implies that the effect of
magnetic excitations on neighboring sites increases with decreasing value
of $\Gamma$.
As regards $f_{\nu}^{(2)}$'s, $|f_2^{(2)}|$
$\sim 2 |f_3^{(2)}|$ $\gg |f_4^{(2)}|$ for
$\Gamma/J_{\perp}=1.0 \sim 2.4$.
This suggests that the exchange type
disturbance does not extend beyond the nearest
neighbor site in both leg and rung directions.

\subsection{RIXS spectra}
\begin{figure}
\includegraphics[width=8.0cm]{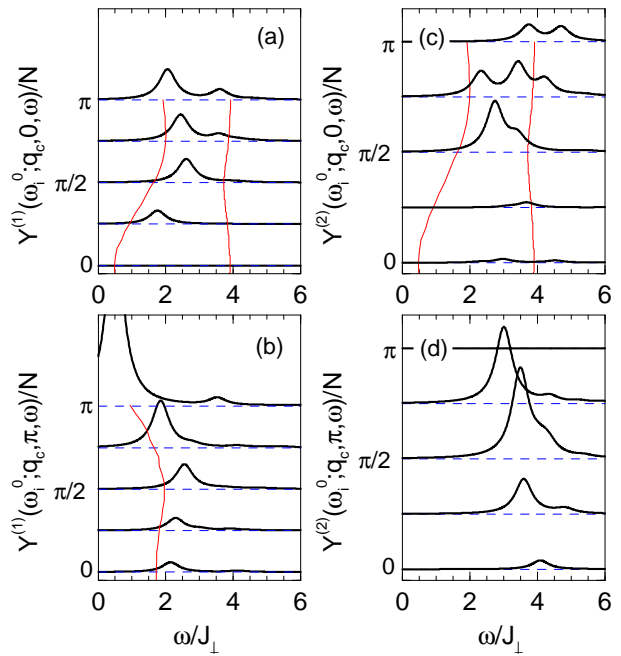}%
\caption{\label{fig.spectrum} (Color online)
Spin correlation functions calculated on a 
$2 \times 8$ spin ladder as a function of the energy
loss $\omega$ for available $q_c$ values.
Left panel is one-spin
correlation function with polarization changing channel
and right panel is two-spin correlation function
with polarization
conserving channel.
Lower and upper panels are for $q_a =0$ and $\pi$,
respectively. The (red) thin lines in panels (a) and (c)
represent the lower and upper boundaries expected
from the two-triplon continuum.\cite{Schlappa2009}
The (red) thin line in panel (b) is the dispersion
curve expected from the one-triplon 
excitation.\cite{Schlappa2009}
}
\end{figure}

We calculate the correlation functions 
from Eqs. (\ref{eq.y1}) and (\ref{eq.y2}).
Figure \ref{fig.spectrum} shows 
$Y^{(1)}(\omega_{i}^{0};q_c,q_a,\omega)$ 
and $Y^{(2)}(\omega_{i}^{0};q_c,q_a,\omega)$ 
numerically calculated with $\Gamma/J_{\perp}=2.4$.
It should be reminded that
$Y^{(1)}(\omega_{i}^{0};q_c,0,\omega)$
and $Y^{(1)}(\omega_{i}^{0};q_c,\pi,\omega)$
include the contributions from
two- and one-triplon excitations, respectively.

In Fig. \ref{fig.spectrum} (a), the energy profile of
$Y^{(1)}(\omega_i^0;q_c,0,\omega)$ is plotted
for accessible values of $q_c$.
The lower and upper boundaries of 
the two-triplon continuum,
shown in Schlappa \textit{et al}.'s paper,\cite{Schlappa2009} 
are also illustrated.
The peak position of 
$Y^{(1)}(\omega_i^0;q_c,0,\omega)$ captures well 
the dispersive behavior of the lower boundary.
Quantitative discrepancy can be ascribed to the
finite size effect.
Since Eq. (\ref{eq.y1}) gives no finite intensity
at the zone center $q_c=0$, our theory, 
unfortunately, cannot
observe the spin gap directly.
However, a little larger cluster may give
the better extrapolation of the magnitude
of the spin gap at the zone center.

In Fig. \ref{fig.spectrum} (b), we can see that the $q_c$ 
dependence of the peak position of 
$Y^{(1)}(\omega_i^0;q_c,\pi,\omega)$
traces one-triplon dispersion relation.\cite{Schlappa2009}
Although Schlappa \textit{et al}. fixed their
experimental configuration such that the
scattering vector was confined within the $bc$-plane, 
RIXS spectra can observe the one-triplon dispersion
around the zone center $q_c \sim 0$ if the projected
momentum transfer $q_a$ could be set as $\pi$.
Such a possibility shall be discussed
in Sec. \ref{sect.4}.

The spectral shape of 
$Y^{(2)}(\omega_{i}^{0};q_c,q_a,\omega)$ as a function of
$\omega$ is shown in Figs. \ref{fig.spectrum} (c) and (d).
It seems to have more weights at higher 
$\omega$ than those of
$Y^{(1)}(\omega_{i}^{0};q_c,q_a,\omega)$.
This tendency is the same as observed in the
analysis of one-dimensional 
system.\cite{Igarashi2012.1D}
The peak intensity of 
$Y^{(2)}(\omega_i^0;q_c,0,\omega)$
is very small when $q_c=0$ $\sim \pi/4$,
but grows rapidly when $q_c$ reaches $\pi/2$.
Then, the spectrum becomes broader and the 
spectral weight shifts to the higher energy
region as $q_c$ goes to the zone boundary.
The peak intensity of 
$Y^{(2)}(\omega_i^0;q_c,\pi,\omega)$
is small at $q_c=0$, peaks at $q_c= \pi/2$,
and vanishes at the zone boundary.

\begin{figure}
\includegraphics[width=8.0cm]{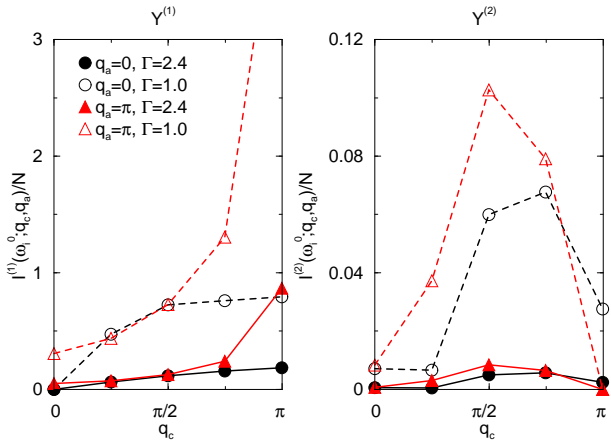}%
\caption{\label{fig.intensity} (Color online)
Frequency-integrated intensities of the correlation
functions (a) $I^{(1)}(\omega_i^0;q_c,q_a)/N$
and (b) $I^{(2)}(\omega_i^0;q_c,q_a)/N$ calculated on
a $2 \times 8$ spin ladder as a function of $q_c$.
(Black) circles and (red) triangles are intensities
for $q_a=0$ and $q_a=\pi$, respectively.
Filled and open symbols correspond to $\Gamma/J_{\perp}
=2.4$ and $1.0$, respectively.
Solid and broken lines are guides to the eye.
}
\end{figure}

Figure \ref{fig.intensity} shows the integrated intensities defined by
\begin{eqnarray}
 I^{(1)}(\omega_i^{0};q_c,q_a) 
&=& \int Y^{(1)}(\omega_i^{0};q_c,q_a,\omega)
 \frac{{\rm d}\omega}{2\pi}, \\
 I^{(2)}(\omega_i^{0};q_c,q_a) 
&=& \int Y^{(2)}(\omega_i^{0};q_c,q_a,\omega)
 \frac{{\rm d}\omega}{2\pi}.
\end{eqnarray}
The $I^{(1)}(\omega_{i}^{0};q_c,0)$ 
vanishes with $q_c\to 0$, and increases gradually with 
$q_c\to\pi$.
The $I^{(1)}(\omega_{i}^{0};q_c,\pi)$, on the other hand,
remains finite at $q_c=0$, and increases rapidly 
with $q_c\to\pi$.
The presence of $f_{2}^{(1)}(\omega_i)$ and $f_{4}^{(1)}(\omega_i)$ 
makes the $q_c$-dependence deviate from that of the 
dynamical structure factor
predicted by the fast collision approximation.
The deviation is small for $\Gamma/J_{\perp}=2.4$, 
but becomes conspicuous with 
$\Gamma/J_{\perp}=1.0$, because of the increase of 
$|f_{2}^{(1)}(\omega_i^0)|$ and $|f_{4}^{(1)}(\omega_i^0)|$.
The $I^{(2)}(\omega_i^0;q_c,0)$ is very small
but finite at $q_c=0$, increases as $q_c$ grows
peaking around $q_c=3\pi/4$,
then remains finite at the zone boundary.
The $I^{(2)}(\omega_i^0;q_c,\pi)$ starts
finite at $q_c=0$,
peaks around $q_c=\pi/2$,
then vanishes at the zone boundary.
The $I^{(2)}(\omega_i^{0};q_c,q_a)$ is found one order of magnitude smaller than
$I^{(1)}(\omega_i^{0};q_c,q_a)$ over the entire
Brillouin zone.

\section{Application to S\lowercase{r}$_{14}$C\lowercase{u}$_{24}$O$_{41}$\label{sect.4}}

\begin{figure}
\includegraphics[width=8.0cm]{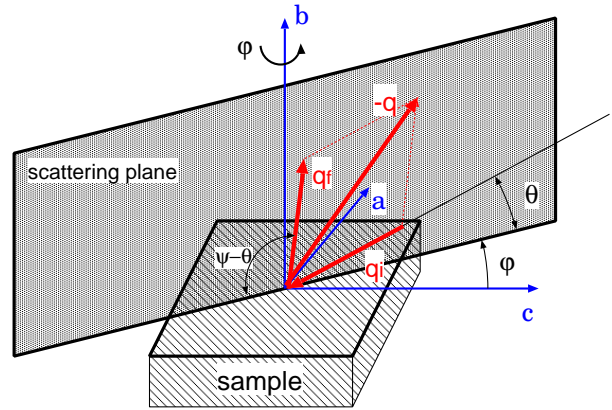}%
\caption{\label{fig.geom} (Color online)
A schematic diagram of the RIXS experimental configuration.
The incident and scattering angles are
$\theta$ and $\psi$, respectively.
The scattering plane is 
perpendicular to the $ca$-plane.
The angle between the scattering plane and the $c$-axis
is $\varphi$, which is fixed to zero in 
the actual experiment.\cite{Schlappa2009}
Scattering vector is defined as 
$-\textbf{q}
=\textbf{q}_f- \textbf{q}_i$.
}
\end{figure}

\begin{figure}
\includegraphics[width=8.0cm]{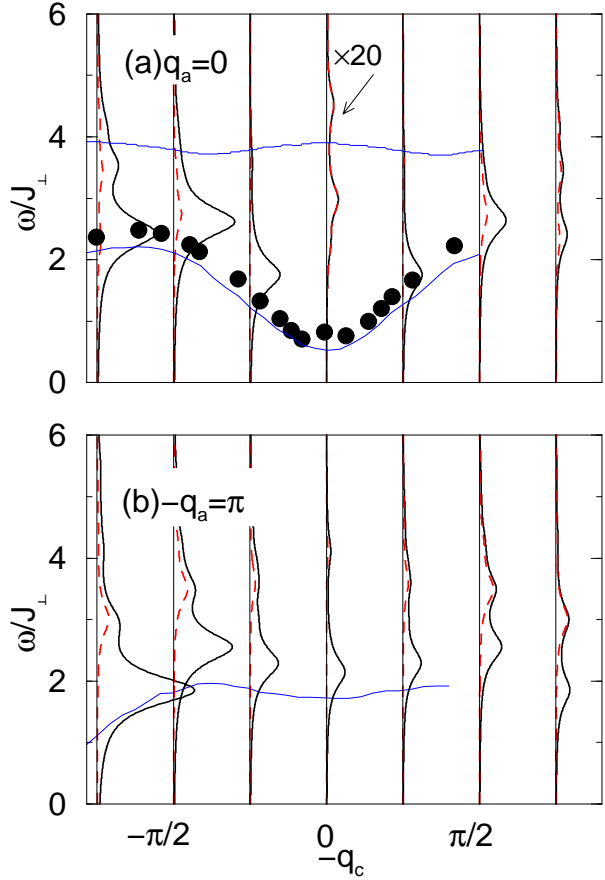}%
\caption{\label{fig.RIXS} (Color online)
RIXS spectra as a function of the energy loss $\omega$
for the available momentum transfer projected on
the $c$-axis ($q_c$) in the $\sigma$ polarization.
Solid (black) and long-dotted (red) 
lines show the total intensity and the contributions
from $Y^{(2)}(\omega_i^0;q_c,q_a,\omega)$, respectively. 
The calculated curves are convoluted with the
Lorentz function with the half-width of half-maximum 78 meV.
(a) For $q_a=0$. Filled circles are the experimental data.\cite{Schlappa2009}
Thin (blue) curves are the lower and upper
boundaries expected from two-triplon 
continuum.\cite{Schlappa2009}
(b) For $-q_a = \pi$. 
Thin (blue) line is the dispersion curve expected from 
one-triplon excitation.\cite{Schlappa2009}
}
\end{figure}

Now we attempt to compare our results
with those observed by the RIXS experiment in
two-leg spin ladder system 
Sr$_{14}$Cu$_{24}$O$_{41}$.\cite{Schlappa2009}
The legs and rungs of the ladders are along the
$c$ and $a$ axes with lattice constants
$a=11.459 \textrm{\AA}$, 
$b=13.368 \textrm{\AA}$,
and along the $c$-direction the unit cell for the 
ladders is $c_L=3.931 \textrm{\AA}$.\cite{McCarron1988}
The lattice parameter is $a_r=3.84 \textrm{\AA}$ along the 
$a$-directions for the ladders.\cite{Notbohm2007}

\subsection{$L_{3}$-edge spectrum}
In this section, we compare our results
with the experiment observed at the
Cu $L_{3}$-edge in Sr$_{14}$Cu$_{24}$O$_{41}$.
Figure \ref{fig.geom} illustrates a schematic
sketch of the experimental 
geometry performed by Schlappa
\textit{et al}.\cite{Schlappa2009} 
Notice that the direction of the scattering vector
in the experiment is opposite to that of ours.
In the following, when the finite numerical values of
the momentum transfer $q_c$ and $q_a$ are mentioned, 
we adopt the experimental definition.
The surface of the sample is perpendicular to
the $b$-axis.  
Since the scattering plane is fixed parallel to 
the $bc(zx)$-plane, the momentum transfer
along the rung direction is zero ($q_a=0$).
Hence, the experiment probed
$Y^{(1)}(\omega_i;q_c,0,\omega)$ and
$Y^{(2)}(\omega_i;q_c,0,\omega)$.
The incident and scattering angles are
$\theta$ and $\psi$, respectively.
In the experiment, both $\psi=90^{\circ}$
and $130^{\circ}$ were used.  We choose the
latter since it covers approximately 90 \% of the Brillouin
zone along the $c$-direction at the Cu $L_3$-edge.
By changing $\theta$, the RIXS spectra for
different $q_c$ were obtained.
The polarization vector of the incident photon is then expressed in the $xyz$ coordinate as 
$\mbox{\boldmath{$\alpha$}}_i=(0,-1,0)$ for the $\sigma$ polarization and
$\mbox{\boldmath{$\alpha$}}_i=(\chi_i^{\pi},0,\tilde{\chi}_i^{\pi})$ 
for the $\pi$ polarization.
Similarly, the polarization of the scattered photon is expressed as
$\mbox{\boldmath{$\alpha$}}_f=(0,-1,0)$ for the $\sigma'$ polarization 
and
$\mbox{\boldmath{$\alpha$}}_f=(\chi_f^{\pi},0,\tilde{\chi}_f^{\pi})$ 
for the $\pi'$ polarization. 
The polarization is usually separated with the incident photon, 
but not separated with the scattered photon in experiments. 
In such a situation, we may express the RIXS spectra depending on 
the polarization of the incident photon as
\begin{eqnarray}
&& I(\omega_i;q_c,q_a,\omega) \nonumber \\
&\propto&
 \left[\left(\frac{\chi_f^{\pi}}{2}\right)^2 
       Y^{(1)}(\omega_i;q_c,q_a,\omega)
     + Y^{(2)}(\omega_i;q_c,q_a,\omega)\right],
 \nonumber \\
\label{eq.spec.final.sig}
\end{eqnarray}
for $\sigma$-polarization, and 
\begin{eqnarray}
&& I(\omega_i;q_c,q_a,\omega) \nonumber \\
&\propto&
 \left[\left(\frac{\chi_i^{\pi}}{2}\right)^2
  Y^{(1)}(\omega_i;q_c,q_a,\omega)
     +(\chi_f^{\pi}\chi_i^{\pi})^2
  Y^{(2)}(\omega_i;q_c,q_a,\omega)\right],
 \nonumber \\
\label{eq.spec.final.pi}
\end{eqnarray}
for $\pi$-polarization.
The contribution of $Y^{(2)}(\omega_i;q_c,q_a,\omega)$ 
relative to that of $Y^{(1)}(\omega_i;q_c,q_a,\omega)$ 
is enhanced by
$( 2/\chi_{f}^{\pi})^2$ in the $\sigma$ polarization. The contribution of
$Y^{(2)}(\omega_i;q_c,q_a,\omega)$ in the $\pi$
polarization is smaller than that in the $\sigma$ polarization by a factor
$(\chi_{f}^{\pi}\chi_{i}^{\pi})^2$.

Figures \ref{fig.RIXS} (a) and (b) show the RIXS spectra 
as a function of energy loss $\omega$ with $-q_a=0$ 
and $\pi$, respectively, in the $\sigma$ polarization.
We put $\Gamma/J_{\perp}=2.4$.
The calculated curves are convoluted by the Lorentzian function with
the half-width-half-maximum of the possible resolution, 
$78$ meV.
Since the contribution from 
$Y^{(1)}(\omega_i;q_c,q_a,\omega)$ is
larger than that from 
$Y^{(2)}(\omega_i;q_c,q_a,\omega)$ in most
conditions, the dispersive behavior of the
lower boundary of two-triplon continuum as well
as one-triplon dispersion can be well reproduced
as is shown in Fig. \ref{fig.spectrum}.

Since the experiment has measured $q_a=0$ mode alone,
the one-triplon dispersion explained above has not
detected yet.
Here, let us discuss a possible configuration to 
access $-q_a=\pi$ 
mode by rotating sample by angle $\varphi$ around the 
$b$-axis (see Fig. \ref{fig.geom}).
An easy way to access $-q_a=\pi$ mode is
setting $\varphi=90^{\circ}$.
Then, we can see that when 
$\psi \geq 138^{\circ}$ at the Cu $L_3$-edge, 
the incident angle $\theta$ becomes positive, 
i.e., such a configuration is mathematically allowed.
Examples are 
$\theta \simeq 2.5^{\circ}$ for $\psi=140^{\circ}$, and
$\theta \simeq 9^{\circ}$ for $\psi=150^{\circ}$.
As seen from Figs. \ref{fig.RIXS} (a) and
(b), we find that the intensity expected from 
$-q_a=\pi$ mode is roughly the same order of magnitude observed in the experiment for $q_a=0$.
Therefore, we assert that 
local minimum of the one-triplon 
dispersion at $-\textbf{q}=$$-(q_c,q_a)$
$=$$(0,\pi)$ can be observed by the RIXS spectrum.
The larger $\psi$ can be,
the larger attainable maximum value 
of $q_c$ becomes away from the zone center.

Next, we concentrate on the contribution of
$Y^{(2)}(\omega_i^0;q_c,q_a,\omega)$, which
is smaller than that of 
$Y^{(1)}(\omega_i^0;q_c,q_a,\omega)$
even in the presence of the geometrical coefficients
appeared in Eqs. (\ref{eq.spec.final.sig}) 
and (\ref{eq.spec.final.pi}).
Though it is small, the contribution of
$Y^{(2)}(\omega_i^0;q_c,q_a,\omega)$ has weight 
in the higher energy transfer region.
The intensity grows larger as $|q_c|$ increases
with $|q_c| \rightarrow \pi/2$, then decreases  
toward the zone boundary.
In Schlappa \textit{et al}.'s data, for each $q_c$,
the intensity is accumulated within the region centered
at the peak position with half-width about $0.1$ eV.
However, 
around $-q_c = -0.8\pi \sim -0.6\pi$,
the tail of the intensity extends 
to the higher energy side about 0.2 eV.
This trend coincides with that shown by the two-spin
correlation function, which leads us to speculate
that the observed weak intensities in the high energy 
region may be 
attributed to the contribution of
$Y^{(2)}(\omega_i^0;q_c,q_a,\omega)$.

\begin{figure}
\includegraphics[width=8.0cm]{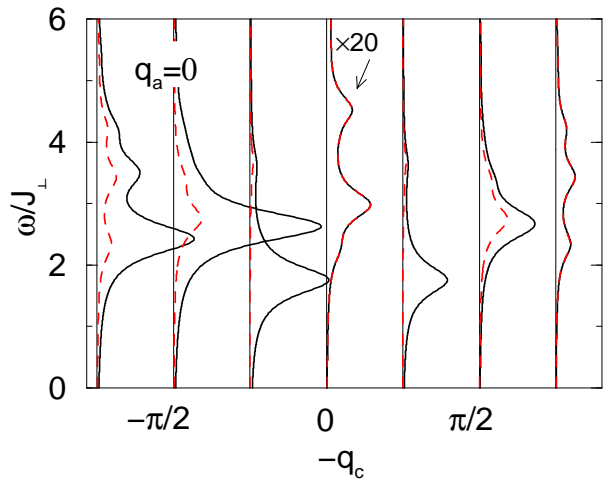}%
\caption{\label{fig.M3edge} (Color online)
RIXS spectra at the Cu $M_{3}$-edge as a function of the energy loss $\omega$
for the available momentum transfer projected on
the $c$-axis ($q_c$) in the $\sigma$ polarization.
Solid (black) and long-dotted (red) 
lines show the total intensity and the contributions
from $Y^{(2)}(\omega_i^0;q_c,q_a,\omega)$, respectively. 
The calculated curves are convoluted with the
Lorentz function with the half-width of half-maximum 78 meV.
}
\end{figure}

\subsection{$M_{3}$-edge spectrum}
Our analysis has shown that the contribution of
$Y^{(2)}(\omega_i^0;q_c,q_a,\omega)$ plays a minor
role in Sr$_{14}$Cu$_{24}$O$_{41}$ in
making a quantitative description of the $L$-edge processes.
On the other hand, we have confirmed that 
when the ratio $\Gamma/J_{\perp}$
becomes smaller, the contribution of
$Y^{(2)}(\omega_i^0;q_c,q_a,\omega)$ becomes larger.
Therefore, we assert that the RIXS signal 
at the Cu $M_{2,3}$-edge may contain significant contribution
of $Y^{(2)}(\omega_i^0;q_c,q_a,\omega)$.
Let us study such a situation and examine the possibility.

The present theory is applicable to analyze the
magnetic excitation spectrum at the Cu $M_{2,3}$-edge.
In the following, we focus on the $M_{3}$-edge case.
A prominent difference between the $M_{3}$-edge and 
$L_{3}$-edge events appears in the values of parameters.
Since the core-hole is in the $3p$ instead of $2p$ state,
the resonant energy at the $M_{3}$-edge (747 eV)
becomes smaller than that at the $L_{3}$-edge (931 eV).
Thus, the core-hole lime-time broadening width
becomes smaller value, for instance, 
$\Gamma=0.2$ eV,\cite{Wray2012}
which results in $\Gamma/J_{\perp} =1.6$.
The smaller resonant energy leads us to the different value of $|\textbf{q}_i|$.
This makes geometrical condition the equipment should
satisfy tougher, but attainable.
Figure \ref{fig.M3edge} shows the calculated result
of the $M_{3}$-edge spectrum 
for $q_a=0$ with $\psi=130^{\circ}$ and $\sigma$ incident
polarization. In this case, the transferred momentum
projected along the leg ($q_c$) can be covered up to 
nearly $80 \%$ of the first Brillouin zone.
The contribution 
of the two-spin correlation function to the total spectrum
at the $M_3$-edge is 
larger than that at the $L_3$-edge.
However, it is hard to distinguish pure 
$Y^{(2)}(\omega_i^0;q_c,0,\omega)$ contribution
since it is hidden by the large 
$Y^{(1)}(\omega_i^0;q_c,0,\omega)$ contribution 
for most $q_c$.
An important difference is found at $-q_c=0.75 \pi$
where the total intensity is almost exclusively generated
by $Y^{(2)}(\omega_i^0;q_c,0,\omega)$.
This value of $q_c$ is near the geometrical
limit but within the reach.
Experiments have been reported in other Cu-oxides
such as CaCuO$_{2}$ and 
SrCuO$_{2}$ at the Cu 
$M_{2,3}$-edges.\cite{Freelon2008,Wray2012}
We hope the RIXS measurement of at Cu $M_{2,3}$-edge
in Sr$_{14}$Cu$_{24}$O$_{41}$
will be carried out in near future. 

\section{Summary and discussion\label{sect.5}}

We have studied the magnetic excitation spectra 
of the $L$-edge RIXS in undoped cuprate, in particular,
in quasi one-dimensional two-leg spin ladder system.
We have analyzed the second-order dipole allowed process 
through the intermediate state, 
in which there is no spin degree of freedom at the 
core-hole site.  This nature of the intermediate state 
is found to affect strongly the transition amplitudes of 
spin excitations.
Then, the RIXS spectra have been derived as 
the one-spin and two-spin 
correlation functions in the channels with and 
without changing polarization, respectively. 
Note that they include
the contributions of the magnetic excitations 
not only on the core-hole site 
but also on the neighboring sites.
The correlation functions are expressed by 
a set of numerical coefficients reflecting 
the weight of each magnetic excitations.

Once the expressions of the correlation functions are 
obtained, we could concentrate on
evaluating them with approximation methods
available.
The coefficients could be rather accurately 
obtained numerically in a system with small size,
since the relevant excitations are restricted around the
core-hole site.
We have evaluated them and, subsequently, the
correlation functions in a finite-size
two-leg ladder.
In the two-leg ladder system, 
the one-spin correlation function
includes both one- and two-triplon excitations with $S=1$.
We find one- and two-triplon excitations 
emerge separately by choosing rung wave number 
$-q_a=\pi$ and $0$, respectively.
An application to Sr$_{14}$Cu$_{24}$O$_{41}$
has revealed that the calculated RIXS spectrum 
captures well the dispersive behavior shown by 
the lower boundary of two-triplon continuum. 
By adjusting the geometrical 
configuration of the experiment, one-triplon
dispersion around the zone center could be 
detectable in the RIXS measurement.
The observed weak intensity in the higher 
energy region around $-q_c = -0.8\pi \sim -0.6\pi$
might be the contribution from 
the two-spin correlation function, which could be clearly
detected at the Cu $M_{3}$-edge. 

We finally comment on the plausibility to
adopt spin only model to explain the
current experiment. Although our theory
has reproduced semi-quantitatively well
the lower boundary of the two-triplon 
continuum over a wide range of the Brillouin
zone, the corresponding spin correlation
function has no intensity at the zone 
center. On the other hand, a small cluster
analysis on the Hubbard model presented by 
Schlappa \textit{et al}. has given a 
finite intensity at $q_c=0$.
Since the RIXS processes in the actual 
materials are very complicated, there is a
possibility that the observed RIXS signals
involves something missing in the spin correlation
function.
For the $K$-edge RIXS spectra, for instance,
Jia \textit{et al}.
has demonstrated that the RIXS intensity and
the spin dynamical structure factor show
difference in several systems.\cite{Jia2011}
Whether the same is true to the present
system is an intriguing problem and relegated to
a future study.  

\begin{acknowledgments}
We thank M. Grioni and H. M. R\o nnow for valuable discussions. T. N. thanks to T. Hikihara
for helpful discussion.
This work was partially supported by a Grant-in-Aid for Scientific Research 
from the Ministry of Education, Culture, Sports, Science and Technology
of the Japanese Government.

\end{acknowledgments}

\bibliographystyle{apsrev}
\bibliography{paper}

\begin{thebibliography}{47}
\bibitem[{Com()}]{Com0}
\bibinfo{note}{Notice that the stoichiometric system has six holes. However,
  since they reside mainly on the chain,\cite{Nucker2000} the Cu ion in the
  ladder part has $(3d)^9$ configuration so that the system is called as
  undoped.}
\end{thebibliography}

\end{document}